# Unravelling the unique kinetic interactions between $N_2O$ and unsaturated hydrocarbons

Hongqing Wu[a,^], Guojie Liang[a,^], Tianzhou Jiang[a], Fan Li[b], Yang Li[d], Rongpei Jiang[e], Ruoyue Tang[a], Song Cheng[a,c]*

[a] *Department of Mechanical Engineering, The Hong Kong Polytechnic University, Kowloon, Hong Kong SAR 999077, China*
[b] *Department of Aeronautical and Aviation Engineering, The Hong Kong Polytechnic University, Kowloon, Hong Kong SAR 999077, China*
[c] *Research Institute for Smart Energy, The Hong Kong Polytechnic University, Kowloon, Hong Kong SAR 999077, China*
[d] *National Key Laboratory of Solid Propulsion, School of Astronautics, Northwestern Polytechnical University, Xi'an, China*
[e] *Beijing Institute of Aerospace Testing Technology, Beijing, China*

^ Both authors contribute equally to this paper.
* Corresponding authors: Song Cheng
Phone: +852 2766 6668
Email: songcheng@polyu.edu.hk

**Abstract:**

The interaction between unsaturated hydrocarbons and $N_2O$ has attracted considerable attention in recent years due to their important roles as potential propellants for advanced propulsion systems (e.g. NOFBX), key combustion intermediates in EGR systems, and as major pollutants and precursors in atmospheric chemistry. Although experimental studies and kinetic models have been developed to investigate its fuel chemistry, discrepancies remain between modeled and measured ignition delay times at low temperatures. In this work, we characterize previously unreported direct interaction pathways between $N_2O$ and unsaturated hydrocarbons ($C_2H_4$, $C_3H_6$, $C_2H_2$, and $C_3H_4$) through quantum chemistry calculations, comprehensive kinetic modeling, and experimental validation. These reactions proceed via O-atom addition from $N_2O$ to unsaturated hydrocarbons, forming five-membered ring intermediates that decompose into $N_2$ and hydrocarbon-specific products. Distinct mechanistic differences are identified between alkenes and alkynes, arising from the disparity in N–C bond lengths within the intermediates (~1.480 Å vs. ~1.381 Å), which governs their decomposition pathways. The corresponding rate coefficients are determined and implemented into multiple kinetic models, with autoignition simulations showing a pronounced promoting effect on model reactivity and improved agreement with experiments, especially at low temperatures. Flux analysis further reveals that the new pathways suppress conventional inhibiting channels while enabling aldehyde- and ketone-forming pathways that enhance overall reactivity. This work provides a more complete description of $N_2O$–hydrocarbon interactions, advancing predictive capability for combustion and atmospheric chemistry.

*Keywords: unsaturated hydrocarbons; $N_2O$; kinetic interactions; ab initio calculations; kinetic modeling and validation.*

**Novelty and Significance Statement**

To date, direct kinetic interactions between unsaturated hydrocarbons and $N_2O$ remain unknown and are missing from all existing chemistry models, despite the important roles of unsaturated hydrocarbons and $N_2O$ as potential propellants for advanced propulsion systems (e.g. NOFBX), key combustion intermediates in EGR systems, and as major pollutant precursors in atmospheric chemistry. This study is the first study that addresses these gaps by systematically quantifying the interaction kinetics between $N_2O$ and two types of unsaturated hydrocarbons: alkenes and alkynes, with rate parameters also determined via quantum chemistry computations. New and unique kinetic interactions are first revealed, with similarities and differences quantified across different reaction sites and unsaturated hydrocarbons, which can lead to the formation of either ketones or aldehydes, with both being important derivatives in combustion and atmospheric chemistry. Incorporating these new pathways and rate coefficients into four kinetic models consistently leads to a substantial reduction in predicted ignition delay times at low temperatures, bringing the simulations into closer agreement with experimental data. These findings address the long-standing overprediction of low-temperature IDTs for unsaturated hydrocarbons/$N_2O$ mixtures.

## 1. Introduction

Nitrous oxide ($N_2O$), a major component of the $NO_x$ family, is a reactive gas commonly produced during combustion processes [1]. It is stable and comparatively unreactive at room temperatures [2]. Moreover, as a strong oxidant, it can sustain combustion even in the absence of molecular oxygen [3]. These benefits make $N_2O$ an ideal candidate for rocket propulsion applications. On the other hand, unsaturated hydrocarbons, containing double or triple bonds between adjacent carbon atoms, are highly reactive and readily undergo addition reactions with elemental halogens, hydrogen halides, alcohols, and many other compounds [4-6]. They also serve as potential fuels or propellants for advanced propulsion systems and are key intermediate products in alkane oxidation [7].

The co-existence of $N_2O$ and unsaturated hydrocarbons can be found in many applications. For instance, the need of high-performance propellants for advanced propulsion systems has pushed the development of new propellants based on N₂O and unsaturated hydrocarbons. In the past decades, hydrazine and hydrazine derivatives like monomethyl hydrazine (MMH) and unsymmetrical dimethyl hydrazine (DMH) are used for spacecraft propulsion applications due to their advantages of long term storable, easy to decompose via catalyst, low risk of unwanted decomposition or explosion, etc. [8,9]. However, these propellants are highly toxic and carcinogenic [10], resulting in high handling and transportation costs. In recent years, NOFBX (Nitrous Oxide Fuel Blend) has emerged as a promising alternative due to advantages including non-toxic nature, high performance, and low cost [11-13]. As a monopropellant, NOFBX is a mixture of unsaturated hydrocarbons (e.g. $C_2H_2$, $C_3H_4$, $C_2H_4$, or $C_3H_6$) with nitrous oxide. On the other hand, exhaust gas recirculation (EGR) has also received significant attention due to its ability to mitigate $NO_x$ formation and enhance system efficiency. Extensive research has been devoted to understanding the underlying fundamentals for EGR [14-17]. During EGR operation, the N₂O remaining in the exhaust gas can interact with the unsaturated hydrocarbons produced from fuel decomposition and oxidation, affecting global combustion and emission characteristics of the engine system. In addition, N₂O and unsaturated hydrocarbons have been identified as major precursors for aldehydes or ketones formation in the atmosphere, and they play an important role in the ozone ($O_3$) destruction cycle. [18-22]. These applications make it necessary to understand the chemical kinetic interactions between unsaturated hydrocarbons and N₂O.

There have been only a few experimental studies on the combustion characteristics of unsaturated hydrocarbons with N₂O blending. In the early 1960s, Trenwith [23] studied the oxidation of $C_2H_4$ by N₂O at temperatures of 555−602 °C and pressures of 20−100 mm Hg. $CH_3CHO$ was identified as a primary product from $C_2H_4$ and N₂O reactions, and global reaction rate constants were proposed for consumption of $C_2H_4$ by N₂O. More recently, Deng et al. [24] measured the ignition delay times of stoichiometric $C_2H_4/O_2/N_2O$/Ar mixtures with molar blending ratios of $N_2O/(N_2O + O_2)$ = 0, 50, 80, and 100%. Experiments were conducted in a shock tube at pressures of 1.2−10 atm, equivalence ratios of 0.5−2.0, and temperatures of 1214−1817 K, from which it was found that the ignition delay times of $C_2H_4$ increase as the N₂O concentration increases at a given pressure and temperature. Naumann et al. [8] and Kick et al. [25] investigated the ignition delay times and laminar flame speeds of ethene/ nitrous oxide mixtures diluted with nitrogen, carbon dioxide, or Argon at pressures ranging from 1 to 16 bar and temperatures of 1000–2000 K using a shock tube. Based on their experimental results, they modified several reactions within the nitrogen subsystem to improve the predictive capability of the publicly available GRI 3.0 mechanism. Feng et al. [11] measured the ignition delay times of a $N_2O-C_2H_4$ NOFBX propellant using a rapid compression machine under $T_C$ = 885–940 K, $P_C$ = 2.5–4.3 MPa, and equivalence ratios of 1.05–1.35. Modeling was also conducted using existing chemistry models, where the results showed notable discrepancies between experimental measurements and model predictions at low temperatures, indicating the insufficiency in existing chemistry models for N₂O and $C_2H_4$ interactions.

Although the experimental observations reviewed above have shown strong kinetic interactions between N₂O and unsaturated hydrocarbons and the gasoline chemistries are continuously refined [26-28], theoretical studies of the interactive kinetics between unsaturated hydrocarbons and N₂O remain seriously limited. To date, there have been only a few *ab initio* studies on the potential energy surfaces for interactions between $C_2H_4$ and N₂O, as well as between $C_2H_2$ and N₂O, as critically reviewed in [1], while no studies are found for other unsaturated hydrocarbons. Unfortunately, the existing studies on $C_2H_4$ and N₂O did not determine any rate parameters nor provided the optimized structures. As such, the existing chemistry models for unsaturated hydrocarbon/N₂O systems currently do not include their direct interacting reactions. Parmon et al. [29] proposed that if the C=C bond is located at the terminal of an unsaturated hydrocarbon molecule, the O atom from N₂O can be added to either carbon atom of the double bond, yielding ketones or aldehydes depending on the O atom addition site. Although these reacting pathways can be important at low temperatures due to the low energy barrier for such addition reactions, there have been no studies on this. Previous research from the authors' group [30-32] have revealed important direct interacting reactions between $NO_x$ and unsaturated hydrocarbons (e.g., RH + $NO_2$ pathways, RO˙ + NO pathways) where these reactions are found to greatly affect fuel reactivity. The current study serves as a continuation of this dedicated effort, with the aims to: (a) investigate the direct interacting pathways and, for the first time, their respective rate parameters between N₂O and $C_2$-$C_3$ alkenes and alkynes ($C_2H_4$, $C_3H_6$, $C_2H_2$, and $C_3H_4$); (b) elucidate the mechanistic differences in direct interacting pathways steps between alkene/N₂O and alkyne/N₂O; and (c) systematically evaluate the impact of these reactions on ignition delay time predictions through kinetic modeling.

## 2. Computational methods

### 2.1. Potential energy surfaces

Electronic geometries, vibrational frequencies and zero-point energies for all species involved in the 5 reactions (including reactants, products, complexes, intermediates (INTs), transition states (TS's)) are calculated at the M06-2X method [33] coupled with the 6-311++G(d,p) basis set [34-36]. Following this, conformer search at the same level of theory is conducted to ensure the optimized structures retain the lowest energy. Intrinsic reaction coordinate (IRC) calculations are carried out at the same level of theory to ensure that the transition state connects the respective reactants with the respective product complex. 1-D hindered rotor treatment [37] are also obtained at the M06–2X/6–311++G(d, p) level of theory for the low frequency torsional modes between non-hydrogen atoms in all of the reactants, TS's, complexes, intermediates and products, with a total of 36 scans (i.e., 10 degrees increment in the respective dihedral angle) for each rotor. Scale factors of 0.983 for harmonic frequencies and 0.9698 for ZPEs that were recommended by Zhao and Truhlar [33] are used herein. Single-point energies (SPEs) are further determined for all the species using the CCSD(T) method with the complete basis set limit (CBS) [38] extrapolated by cc-pVTZ and cc-pVQZ [39] as:

$$E_{CBS} = E_{CCSD(T)/cc-pVQZ} + (E_{CCSD(T)/cc-pVQZ} - E_{CCSD(T)/cc-pVTZ}) * \frac{4^4}{5^4 - 4^4} \quad (1)$$

With the CCSD(T) method, attention must be addressed to T1 diagnostic [40] to measure the multi-reference characteristics. The T1 diagnostic values for all the species, as summarized in Table S1 in the Supplementary Material, are below 0.033, which indicates that the SPEs calculated from using single-reference calculation method are reliable in this study. All the calculations mentioned above are performed using ORCA 5.0.4 [41] or GAUSSIAN [42], and the optimized structures for all species, TS's and complexes are summarized in the Supplementary Material.

## 2.2. Rate coefficients calculations

The Master Equation System Solver (MESS) program suite [43] is employed here to calculate the chemical rate coefficients for the reactions via solution of the one-dimensional master equation, based on the chemically significant eigenstate approach of Miller and Klippenstein [44] and the bimolecular species model of Georgievskii and Klippenstein [45]. The frequencies of lower-frequency modes are replaced by the hindered rotor potentials obtained from 1-D scans. Quantum mechanical tunneling corrections assuming the asymmetric Eckart potential (TST/Eck) [46] are applied to obtain the rate coefficient over the temperature range of 298-2000 K. All rate coefficients were fitted to the modified Arrhenius equation, which can be defined as

$$k = AT^n \exp(-E_a/RT) \quad (2)$$

## 2.3. Kinetic modeling

There are four chemistry models which are suitable for predicting the IDTs of unsaturated hydrocarbons/$N_2O$ mixtures, which are: (1) the Aramco model [47], (2) the DLR model [10], (3) the GRI model [48], and (4) the NUIG model [49]. These models are used herein to investigate the impact of the direct interacting reactions between $N_2O$ and unsaturated hydrocarbons on the modeling results. Kinetic modeling of autoignition experiments are conducted using the LLNL-developed fast solver Zero-RK [50].

## 3. Results and discussion

### 3.1. Species and reaction sites

This study investigates the O-atom addition reactions of two alkenes ($C_2H_4$ and $C_3H_6$) and two alkynes ($C_2H_2$ and $C_3H_4$) by nitrous oxide ($N_2O$), resulting in the formation of $N_2$ and their corresponding products. A total of five reaction pathways is explored, as summarized in Fig. 1. Specifically, $C_2H_4$ reacts with $N_2O$ to form $CH_3CHO$ and $N_2$ (Fig. 1(a)). For $C_3H_6$, the oxygen atom of $N_2O$ can be added at two distinct positions: the $v(T)$ site, leading to the formation of $CH_3COCH_3$ (Fig. 1(b)), and the $v(S)$ site, yielding $CH_3CH_2CHO$ (Fig. 1(c)). In the case of alkynes, $C_2H_2$ reacts with $N_2O$ to produce $CH_2CO$ (Fig. 1(d)), while $C_3H_4$ yields $CH_3CHCO$ as the final product (Fig. 1(e)).

The detailed reaction mechanisms and energy profiles for each system will be discussed in the following section.

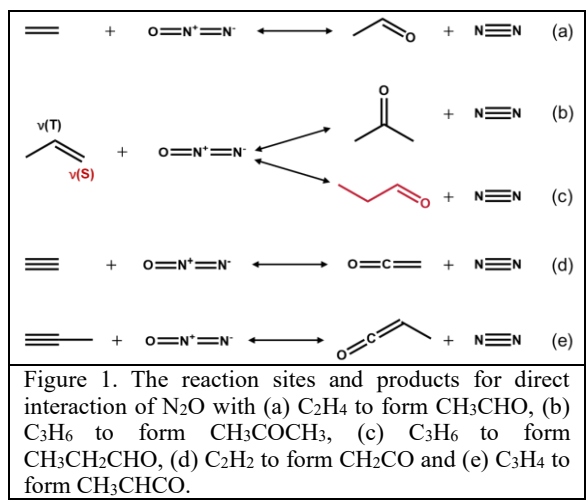

Figure 1. The reaction sites and products for direct interaction of $N_2O$ with (a) $C_2H_4$ to form $CH_3CHO$, (b) $C_3H_6$ to form $CH_3COCH_3$, (c) $C_3H_6$ to form $CH_3CH_2CHO$, (d) $C_2H_2$ to form $CH_2CO$ and (e) $C_3H_4$ to form $CH_3CHCO$.

### 3.2. Reaction path and potential energy surfaces

#### 3.2.1. Alkenes

For alkenes, the reaction pathways leading to the corresponding products are generally similar. Taking $C_2H_4$ as an example, the reaction with $N_2O$ proceeds via the formation of a reactant complex RC_1-1 prior to reaching the transition state TS_1-1, as can be seen in Fig. 2. In the transition state TS_1-1, the nitrogen and oxygen atoms of $N_2O$ approach the two different carbon atoms of $C_2H_4$. Subsequently, both the N and O atoms form bonds with the different carbon atoms, resulting in a five-membered ring intermediate INT_1-1. Following this, the distance between N and O atoms is elongated, leading to the formation of TS_1-2. Thereafter, the N-C bond breaks, accompanied by a hydrogen atom transfer from the -$CH_2O$ moiety to the adjacent carbon atom, ultimately forming a -$CH_3$ group. This step, illustrated in Fig. S1, is particularly interesting as it involves four simultaneous bond rearrangements: cleavage of the N-O and N-C bonds, along with the breaking and formation of two C-H bonds. Finally, the reaction yields $CH_3CHO$ and $N_2$ as products.

For $C_3H_6$, the reaction follows an analogous pathway. However, the specific site of O-atom addition, namely either the $v(S)$ or $v(T)$ site on $C_3H_6$, determines the resulting products. Addition at the $v(S)$ site leads to the formation of $C_2H_5CHO$, while addition at the $v(T)$ site produces $CH_3COCH_3$, as illustrated in Figs. 3 and 4, respectively. The quantum chemistry calculations presented in Figs. 2-4 verify the reaction pathway conjectured by Parmon et al. [29].

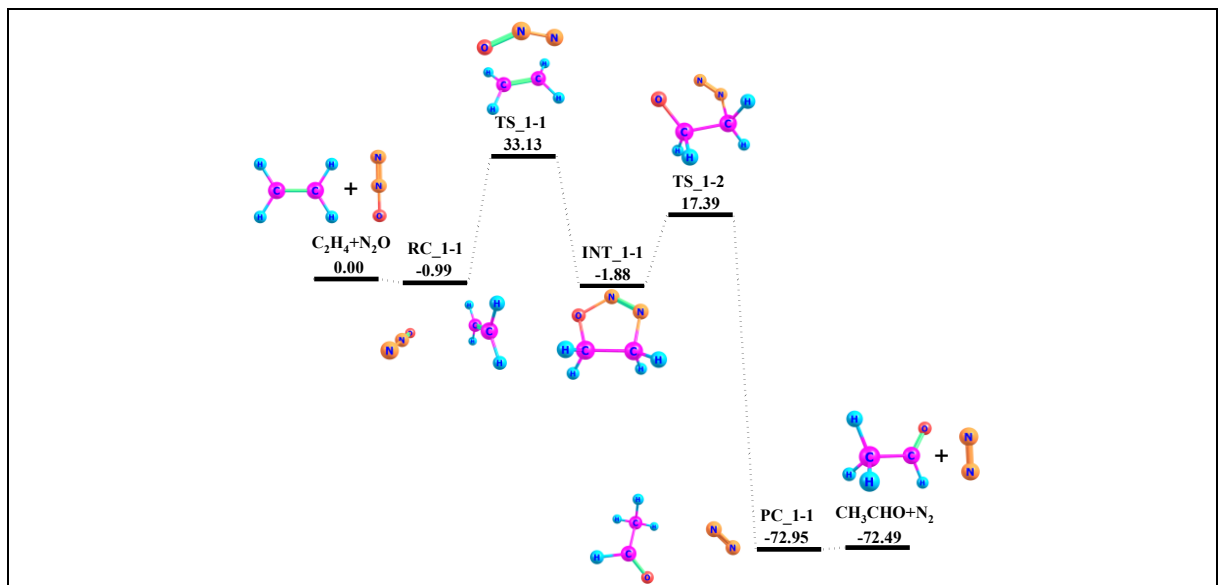

Figure 2. The PES for O-atom addition reaction from $N_2O$ to $C_2H_4$ forming $CH_3CHO$ and $N_2$. All values are in *kcal/mol*.

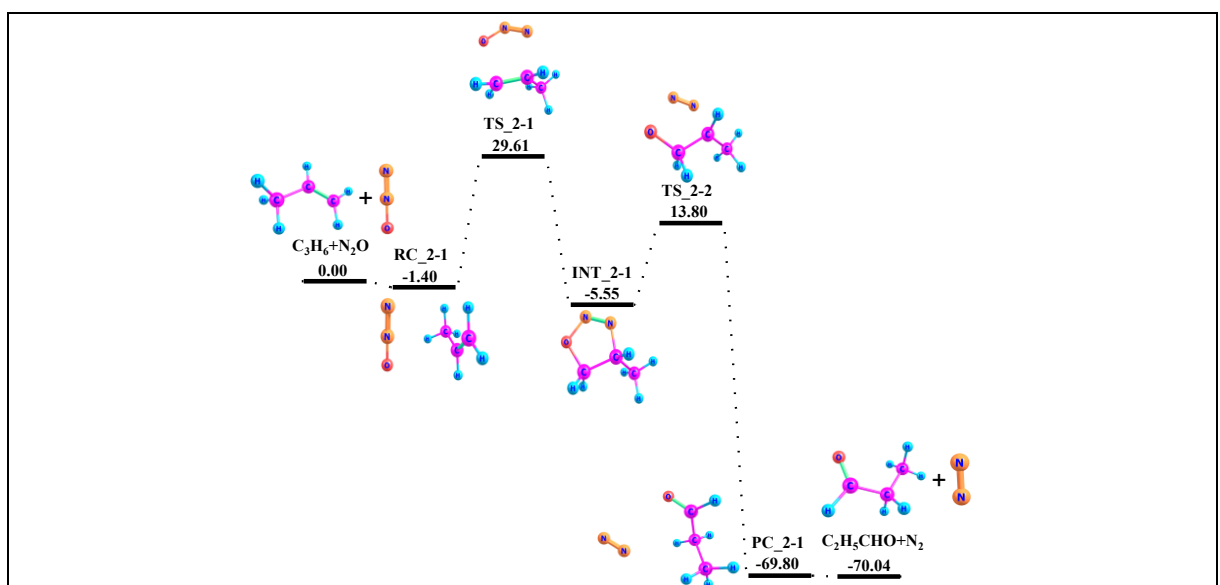

Figure 3. The PES for O-atom addition reaction from $N_2O$ to $C_3H_6$ on the $v(S)$ site, forming $C_2H_5CHO$ and $N_2$. All values are in *kcal/mol*.

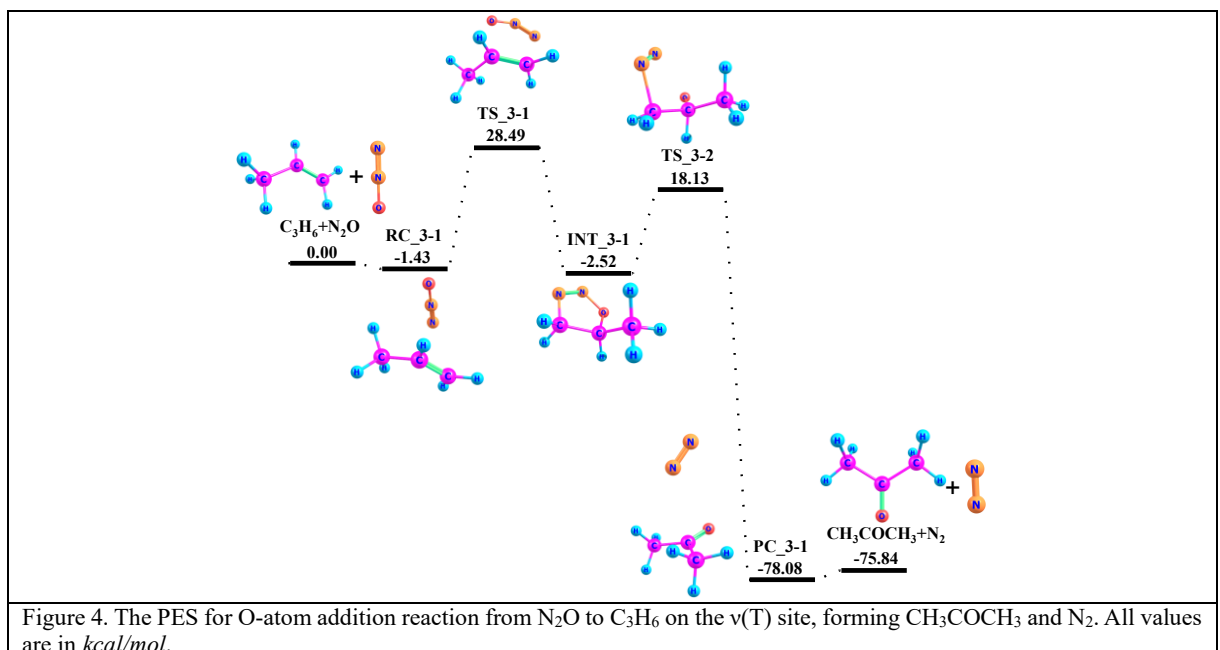

Figure 4. The PES for O-atom addition reaction from N$_2$O to C$_3$H$_6$ on the ν(T) site, forming CH$_3$COCH$_3$ and N$_2$. All values are in *kcal/mol*.

### 3.2.2. Alkynes

For alkynes, the reaction pathways differ significantly from those of alkenes. Taking C$_2$H$_2$ as an example, the N and O atoms of N$_2$O approach the two carbon atoms of C$_2$H$_2$ to form the transition state TS_4-1 via the reactant complex RC_4-1, as can be seen in Fig. 5. In this step, the C≡C triple bond is cleaved to form C=C double bond, and new N-C and O-C bonds are formed, yielding a five-membered ring intermediate INT_4-1. Subsequently, the N-O bond in INT_4-1 breaks via TS_4-2, generating another intermediate, INT_4-2. This step is notably different from the alkene pathways, where both C-N and N-O bonds typically cleave in a single transition state. In contrast, for alkynes, only the N-O bond is cleaved at this stage. Following this, the C-N bond in INT_4-2 is elongated, leading to the formation of transition state TS_4-3. Eventually, the C-N bond breaks, and a hydrogen atom transfer occurs from the -CHO moiety to the adjacent -CH group, resulting in the final products: CH$_2$CO and N$_2$, via the product complex PC_4-1. The detailed reaction pathway from INT_4-2 to PC_4-1 is illustrated in Fig. S2.

For C$_3$H$_4$, although its C≡C triple bond provides two potential sites for O-atom addition, the reaction proceeds preferentially via the terminal carbon atom, ultimately yielding CH$_3$CHCO and N$_2$ as the final products, as shown in Fig. 6. The overall reaction pathway closely resembles that of C$_2$H$_2$ and therefore will not be described in detail here.

### 3.2.3. Validation of determined reaction pathways

To validate the obtained results, this study compares the potential energy surface (PES) and reaction profile for C$_2$H$_2$ with those reported by Karami et al. [51], where B3LYP/6-311++G(3df,3pd) and CCSD(T) levels of theory were used, which differ from the M06-2X/6-311++G(d,p) and CCSD(T)-CBS methods used in this study. Despite the differences in theories, the overall reaction pathway obtained here is consistent with that reported by Karami et al. [51], as can be seen in Fig. 7. Notably, this study includes both the reactant complex (RC) and product complex (PC), which were not identified in [51].

In terms of energy comparison, the calculated energies for TS_4-1, TS_4-3 and the product are in good agreement between the two studies, with discrepancies within 1 kcal/mol. However, a significant difference is observed in the region between INT_4-1 and INT_4-2. The potential energies calculated in this study are consistently lower than those reported by Karami et al. [51], with energy differences of 4.96, 3.70, and 3.74 kcal/mol for INT_4-1, TS_4-1, and INT_4-2, respectively.

To investigate the source of this discrepancy, the geometries of these three key structures are compared, with the results summarized in Fig. 8. For INT_4-1, the largest structural difference lies in the N-O bond length, which is 1.395 Å in this study and 1.453 Å in [51]. For TS_4-2, the N-O bond lengths are 1.910 Å in this study and 1.862 Å in [51], and a significant difference is also found in the ∠NNC bond angle, with 124.3° in this study and 98.1° in [51]. In contrast, the geometries of INT_4-2 in both studies are very similar, with differences in bond lengths and angles within 0.01 Å and 2°, respectively. This study also compares the vibrational frequencies and T1 diagnostic values of the three structures with those reported by Karami et al. [51] in Table 1. The results show only minor differences. Notably, the T1 diagnostic values from this study are lower than those reported by Karami et al. [51], suggesting a more reliable single-reference description.

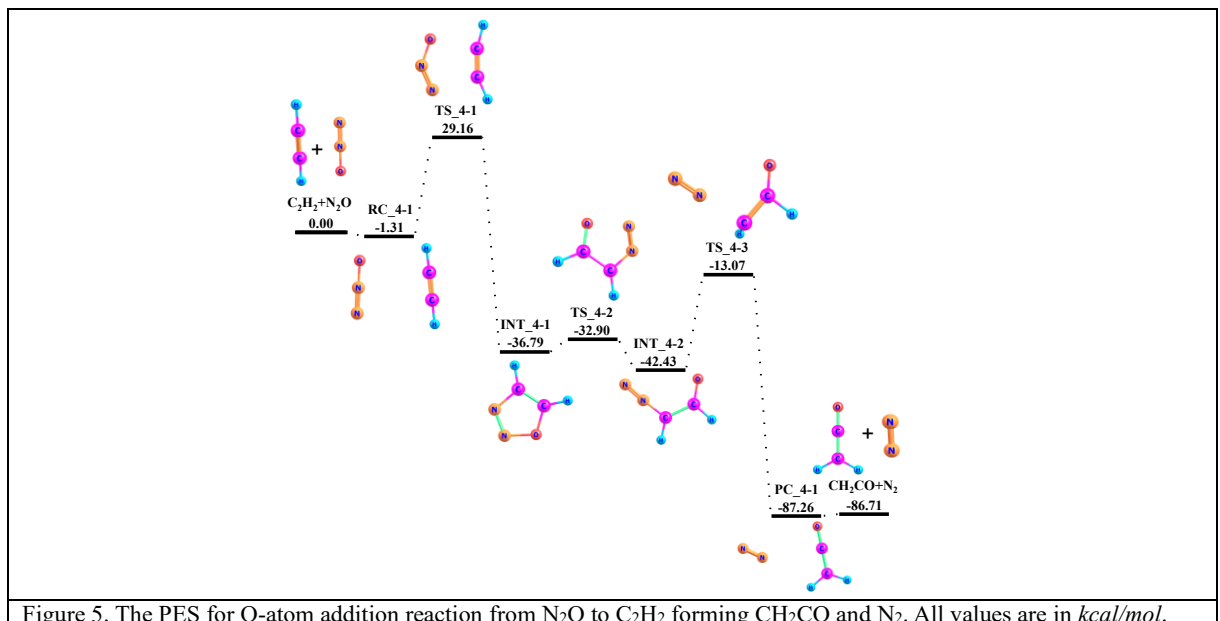

Figure 5. The PES for O-atom addition reaction from $N_2O$ to $C_2H_2$ forming $CH_2CO$ and $N_2$. All values are in *kcal/mol*.

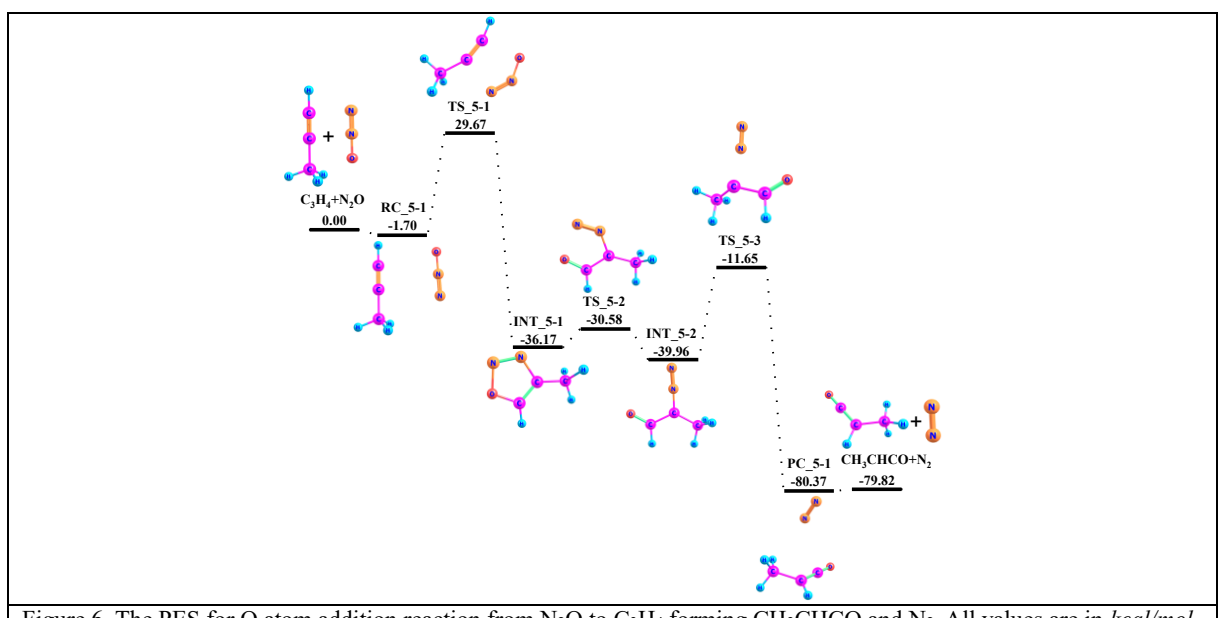

Figure 6. The PES for O atom addition reaction from $N_2O$ to $C_3H_4$ forming $CH_3CHCO$ and $N_2$. All values are in *kcal/mol*.

Table 1. Comparison of the frequencies and T1 diagnostic for INT_4-1, TS_4-2, and INT_4-2, between this study and Karami et al. [51]. All energy values are given in *kcal/mol*.

|  | Frequencies of this study and Karami et al. [51] | T1 diagnostic (this study) | T1 diagnostic (Karami et al. [51]) |
|---|---|---|---|
| INT_4-1 | 629[591], 711[674], 762[677], 821[775], 888[801], 963[940], 1006[969], 1113[1117], 1159[1145], 1177[1183], 1394[1316], 1491[1392], 1599[1534], 3313[3317], 3336[3338] | 0.017 | 0.018 |
| TS_4-2 | -354[-370], 522[473], 598[529], 700[656], 751[705], 875[851], 951[890], 1074[1066], 1133[1095], 1257[1267], 1374[1356], 1555[1467], 1901[2103], 3170[3180], 3350[3358] | 0.028 | 0.029 |
| INT_4-2 | 155[152], 272[247], 444[377], 512[437], 605[535], 816[810], 987[978], 1016[989], 1175[1148], 1378[1391], 1423[1423], 1786[1700], 2308[2360], 3001[3002], 3290[3294] | 0.018 | 0.020 |

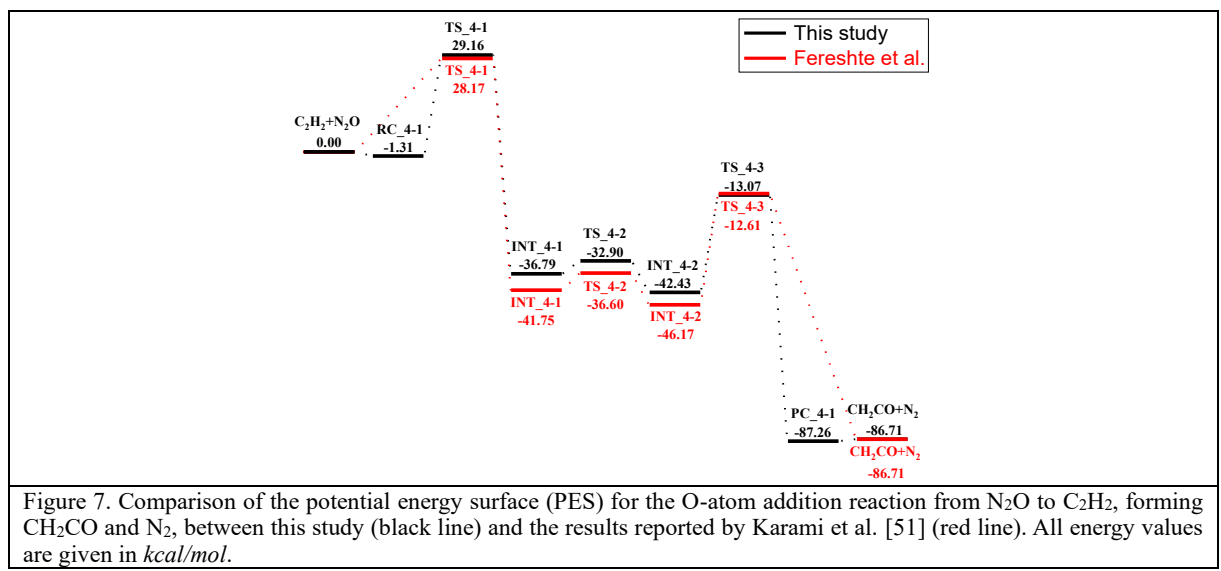

Figure 7. Comparison of the potential energy surface (PES) for the O-atom addition reaction from $N_2O$ to $C_2H_2$, forming $CH_2CO$ and $N_2$, between this study (black line) and the results reported by Karami et al. [51] (red line). All energy values are given in *kcal/mol*.

In conclusion, the discrepancies in geometries, frequencies, and potential energies between the two studies can largely be attributed to differences in the optimization methods—M06–2X/6–311++G(d,p) in this study versus B3LYP/6-311++G(3df,3pd) in [51]. For the single point energy calculation, the specific level of CCSD(T) used by Karami et al. [51] is not reported, which might also contribute to the observed differences. These methodological differences likely account for the deviations seen in the reaction pathway from INT_4-1 to INT_4-2.

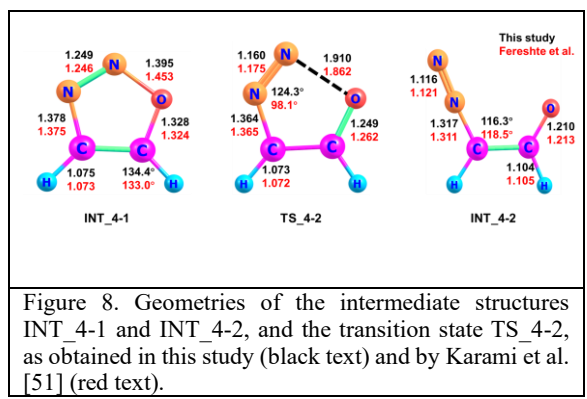

Figure 8. Geometries of the intermediate structures INT_4-1 and INT_4-2, and the transition state TS_4-2, as obtained in this study (black text) and by Karami et al. [51] (red text).

### 3.2.4. The reason for the different reaction steps between alkenes and alkynes

As mentioned above, when alkenes react with $N_2O$, they first form five-membered ring intermediates (INT_X-1). Subsequently, both the N-O bond and the N-C bond break, followed by an H atom shift, ultimately leading to product formation. In contrast, for alkynes, the reaction also begins with the formation of five-membered ring intermediates (INT_X-1), but only the N-O bond breaks at this stage, yielding new intermediates (INT_X-2). Through a distinct sub-reaction, these intermediates eventually convert into the final products. To explore the different reaction steps between alkenes and alkynes, this study further analyzes the geometries of the intermediate structures: INT_X-1 of (a) alkenes and (b) alkynes.

It is evident in Fig. 9 that for all intermediate structures, the lengths of the N-O bonds are very similar, with differences within 0.012 Å, indicating that the energy required to cleave the N-O bond is nearly the same across systems. However, a significant difference lies in the N-C bond length: in alkene intermediates, the N-C bond is approximately 1.480 Å, whereas in alkyne intermediates, it is about 1.381 Å, with a difference of around 0.1 Å. Since bond length is inversely related to bond dissociation energy, a shorter N-C bond generally corresponds to a stronger bond due to greater orbital overlap and increased bond stability. This suggests that breaking the N-C bond in alkyne intermediates requires more energy than in alkene intermediates. Consequently, the five-membered ring intermediates of alkenes can break both the N-C and N-O bonds in a single transition state, while those of alkynes can only cleave the N-O bond at that stage.

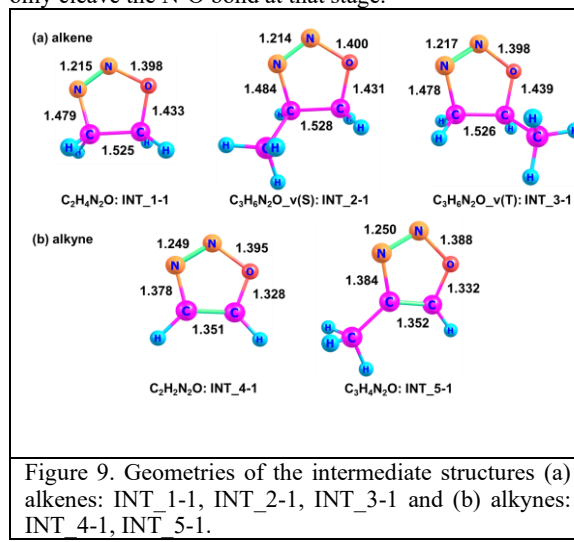

Figure 9. Geometries of the intermediate structures (a) alkenes: INT_1-1, INT_2-1, INT_3-1 and (b) alkynes: INT_4-1, INT_5-1.

### 3.3. Rate coefficients

Table 2. Fit parameters of A, n, and E, for the rate coefficients of O-atom addition from $N_2O$ to alkenes and alkynes.

| Species | O-atom addition site | $A$ (cm$^3$/mol*s) | $n$ | $E$ (cal/mol) |
|---|---|---|---|---|
| $C_2H_4$ | - | 1.033E+04 | 2.217 | 31904.50 |
| $C_3H_6$ | v(S) | 1.359E+04 | 1.955 | 29130.57 |
| $C_3H_6$ | v(T) | 1.150E+07 | 1.176 | 31050.75 |
| $C_2H_2$ | - | 2.122E+05 | 2.014 | 27422.86 |
| $C_3H_4$ | - | 2.808E+04 | 2.069 | 28264.10 |

Based on the calculated potential energy surfaces of all reactions, the corresponding rate coefficients are determined and fitted into Arrhenius parameters, as shown in Fig. 10 and summarized in Table 2, respectively. For alkenes, at low temperatures (e.g. 298 K), the rate coefficient of $C_2H_4$ is lower than those of $C_3H_6$, with the values for different reactive sites of $C_3H_6$ being comparable. Among these, the v(S) site of $C_3H_6$ exhibits the highest rate coefficient. However, as the temperature increases, the rate coefficients of $C_2H_4$ and the v(T) site of $C_3H_6$ surpass that of the v(S) site from $C_3H_6$, which can be attributed to the influence of 1-D hindered rotor effects at elevated temperatures. For alkynes, the rate coefficient of $C_2H_2$ remains higher than that of $C_3H_4$ across the entire temperature range. Moreover, alkynes generally exhibit higher rate coefficients than alkenes, particularly at low temperatures. At 298 K, the rate coefficients for alkynes and alkenes are approximately $10^{-10}$ $cm^3mol^{-1}s^{-1}$ and $10^{-12}$ $cm^3mol^{-1}s^{-1}$, respectively, with a difference of about two orders of magnitude. As temperature increases, this gap narrows to within one order of magnitude at 2000 K.

### 3.4. Model implementation and implications

#### 3.4.1. Ignition delay time

To evaluate the impact of newly introduced reaction pathways on model performance, autoignition simulations are conducted under conditions of 40 bar pressure, an equivalence ratio of 1, and a temperature range of 800-2000 K, using four kinetic mechanisms: (1) the Aramco model [47], (2) the DLR model [10], (3) the GRI model [48], and (4) the NUIG model [49]. Simulation results obtained with the original mechanisms are denoted as 'Original_model name', whereas those obtained with the updated mechanisms incorporating the new reaction pathways are denoted as 'Updated_model name'. A comparative analysis of these results is presented below.

As shown in Fig. 11, the addition of O atoms by $N_2O$ noticeably affects the IDTs of $C_2H_4$, $C_3H_6$, $C_2H_2$, and $C_3H_4$, particularly at low temperatures. At higher temperatures, the ignition delay times (IDT) predicted by both the original and updated models are consistent, and the differences among the four models are negligible. However, at low temperatures, the IDTs predicted by all updated models are consistently shorter than those predicted by the corresponding original models, indicating that the newly introduced pathways enhance the autoignition reactivity of these fuel systems. Notably, for species such as $C_3H_6$, $C_2H_2$, and $C_3H_4$, the updated models are able to predict IDTs at lower temperatures, which is not achievable with the original models. Furthermore, except for $C_3H_4$, the IDTs predicted by the updated models exhibit better agreement with each other than those predicted by the original models, demonstrating improved consistency across different mechanisms. For instance, in the case of $C_2H_2$, as shown in Fig. 11(c), the original Aramco, DLR, and NUIG models can only predict IDTs down to approximately 950 K. In contrast, their updated counterparts can extend predictions down to about 800 K, with the IDTs being up to an order of magnitude shorter. Similarly, the original GRI model can only predict IDTs at temperatures above 1000 K, which is insufficient for capturing the low-temperature chemistry of the fuel system. However, the incorporation of the new reaction pathways addresses this limitation, enabling the updated GRI model to predict IDTs as low as 800 K. This observation suggests that the newly introduced reaction pathways are not only influential but also likely to dominate the low-temperature ignition chemistry.

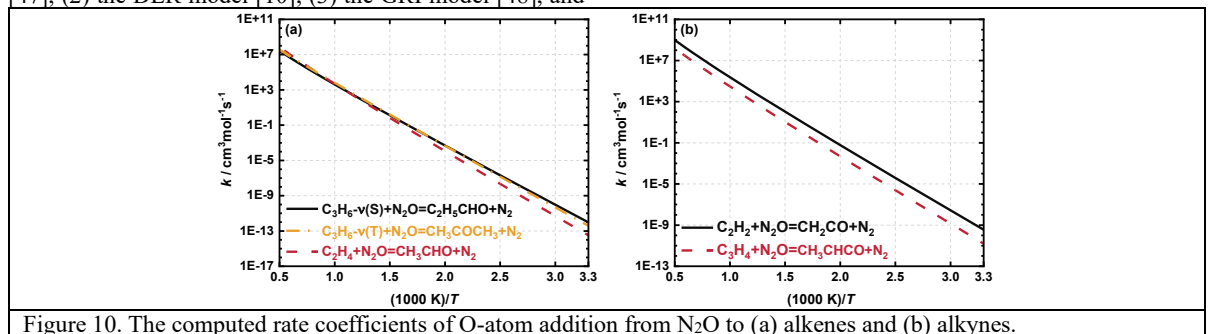

Figure 10. The computed rate coefficients of O-atom addition from $N_2O$ to (a) alkenes and (b) alkynes.

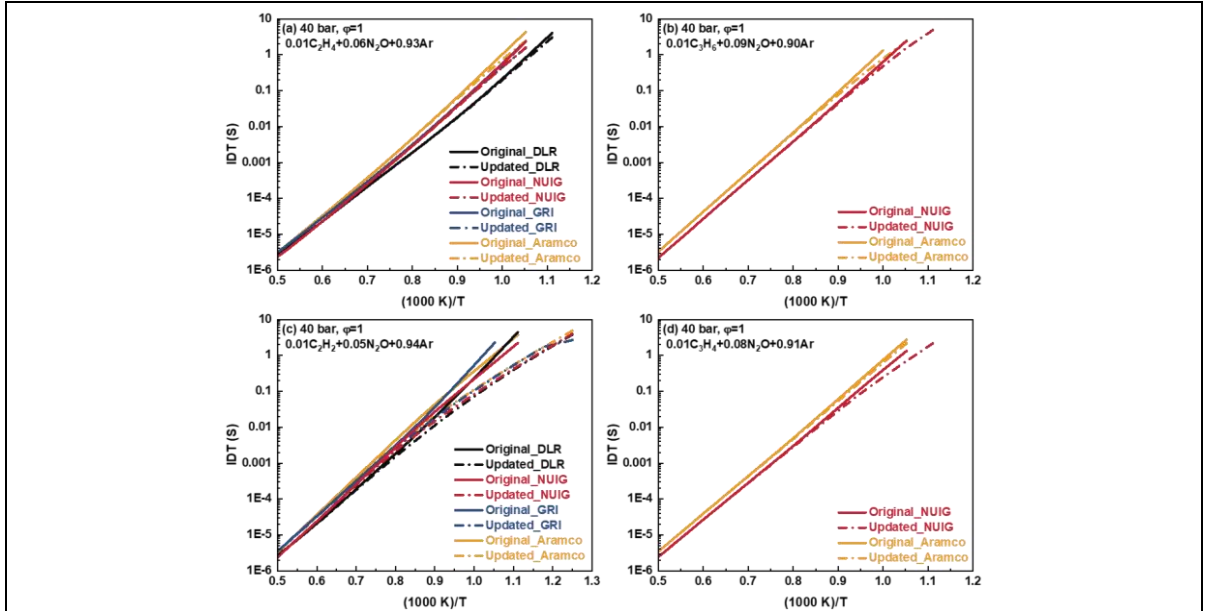

Figure 11. The IDTs of (a) $C_2H_4$, (b) $C_3H_6$, (c) $C_2H_2$, and (d) $C_3H_4$ mixed with $N_2O$ and Ar using both the original and updated chemical kinetic models. Simulations are conducted at pressure of 40 bar, equivalence ratio of 1.0, and temperatures of 800−2000 K.

### 3.4.2. Comparison of ignition delay time between model prediction and experimental results

To verify the impact of the newly introduced reaction pathways, the IDTs predicted by both the original and updated models are further compared with existing experimental data. For the intermediate- and high-temperature ranges, the simulation results are compared with experimental measurements from Naumann et al. [52], conducted using a shock tube, as shown in Fig. 12. The experimental conditions were set at 16 bar, equivalence ratio of 1, and mixture composition of 20% $C_2H_4$ and $N_2O$ diluted in 80% $N_2$. As can be seen in Fig. 12, the IDTs predicted by all four original models show good agreement with the experimental data in the temperature range of 1250-1650 K. However, when temperature is below 1250 K, notable discrepancies emerge between original model predictions and experimental observations. The largest deviation is observed in the original GRI model, which overpredicts the IDT by approximately one order of magnitude. In contrast, the original DLR model yields predictions that are more consistent with the experimental data. For the updated models, the predicted IDTs are uniformly lower than those of their original counterparts at medium temperatures, indicating enhanced alignment with the experimental results. Among all the models, the updated DLR model demonstrates the best overall agreement with the experimental data.

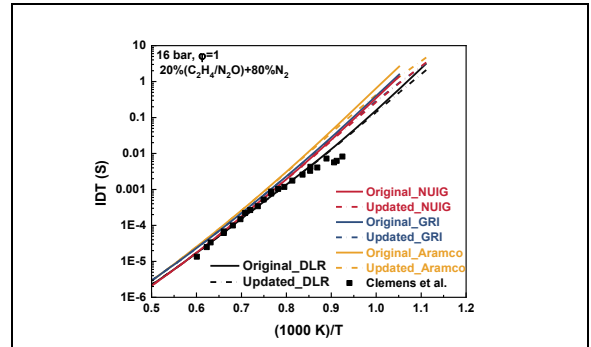

Figure 12. The IDTs of $C_2H_4$ mixed with $N_2O$ and Ar using both the original and updated chemical kinetic models. The experiments were measured by Naumann et al. [52] using a shock tube. Simulations are conducted at pressure of 16 bar, equivalent ratio of 1.0, and temperatures of 800−2000 K.

For low temperature conditions, this study compares the simulated IDTs with the experimental results from Feng et al. [11] that were obtained in a rapid compression machine under the conditions of $T_C$ = 885−940 K, $P_C$ = 3 MPa, $\varphi$ = 1.05 and 1.35 for $N_2O$-$C_2H_4$ propellants. As can be seen from Fig. 13, all original models significantly overpredict the IDTs compared to the experimental data, with discrepancies exceeding one order of magnitude. The limitation in predicting IDTs at low temperatures is observed for all four original models. This is particularly evident with the original Aramco model, which is only capable of predicting IDTs above 950 K and fails to cover the temperature range of the experimental data. By contrast, the updated models extend the predictive capability to lower temperatures and yield consistently shorter IDTs than the original models, resulting in much better agreement with the measurements. Moreover, the IDTs predicted by the updated models are more consistent with one another. This improvement is primarily attributed to the newly introduced reaction pathway, $C_2H_4 + N_2O = CH_3CHO + N_2$, which becomes a dominant pathway in the combustion chemistry of this fuel system at low temperatures.

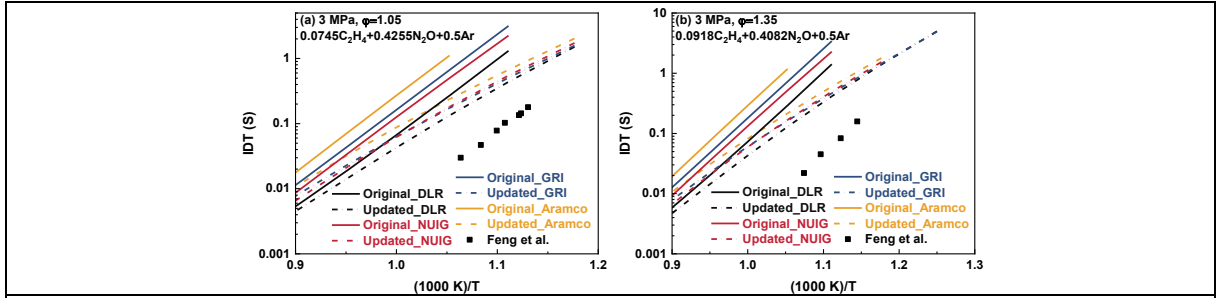

Figure 13. The IDTs of $C_2H_4$ mixed with $N_2O$ and Ar using both the original and updated chemical kinetic models. The experiments were measured by Feng et al. [11] using a rapid compression machine. Simulations are conducted at pressure of 3 MPa, equivalence ratio (φ) of (a) 1.05 and (b) 1.35, and temperatures of 800−1100 K.

### 3.4.3. Sensitivity analysis

To draw further insight into the impact of the O-atom addition into the unsaturated hydrocarbon from $N_2O$ in the fuel systems, brute force sensitivity analysis is conducted on the main ignition delay time using constant volume simulations for the original & updated DLR and NUIG model, due to that the IDTs predicted by both of these models are more consistent with the experimental result. The condition of this analysis is the same as the condition in Section 3.4.1, with temperature fixed at $T_C = 1000$ K. $C_2H_4$ and $C_3H_6$, are selected for analysis. The sensitivity analysis coefficient is defined as:

$$S_{rel} = \ln\left(\frac{\tau^\Delta}{\tau}\right) \Big/ \ln\left(\frac{k^\Delta}{k}\right) \qquad (3)$$

where $\tau^\Delta$ is the main IDT after multiplying the original rate constant by 2, i.e., $k^\Delta = 2 * k$, and $\tau$ is the original ignition delay time. The negative sensitivity coefficient indicates the promotion effect, while the positive sensitivity coefficient indicates the inhibition effect. Therefore, Figures 14 and 15 present the computed sensitivity coefficients of different fuels for the 16 most sensitive reactions.

Figure 14 presents the sensitivity analysis results for $C_2H_4$. For both models, the most sensitive and strongly promoting reaction is the non fuel-specific reaction $N_2O(+M) = N_2 + O(+M)$. In the DLR model, Fig. 14(a), upon introducing the new pathway $C_2H_4 + N_2O = CH_3CHO + N_2$, this reaction emerges as the third most sensitive and promoting reaction. Notably, it also becomes the most sensitive fuel-specific reaction in the system. In addition, the subsequent reaction pathways involving $CH_3CHO$ molecules (e.g., $CH_3CHO + CH_3 = CH_3CO + CH_4$) show increased sensitivity, which further contributes to the overall reactivity of the fuel system. The addition of O atom from $N_2O$ to $C_2H_4$ consistently enhances reactivity by driving the reaction pathway toward $CH_3CHO$ formation and branching. Additionally, the ratio of the most inhibiting fuel-specific reaction (i.e., $C_2H_4 + CH_2 = aC_3H_5 + H$) decreases in the updated model compared with the original model.

In the NUIG model, Fig. 14(b), a similar trend is observed. The new reaction pathway again emerges as the most sensitive fuel-specific reaction, and in this case, it ranks as the second most sensitive reaction overall. Additionally, the sensitivity coefficients of key follow-up reactions, such as $CH_3CHO + H = CH_3CO + H_2$ and $CH_3CHO + CH_3 = CH_3CO + CH_4$, increase significantly, further enhancing the reactivity of the fuel system.

Overall, the pathway $C_2H_4 + N_2O = CH_3CHO + N_2$ and its subsequent reactions play a critical role in enhancing system reactivity to reduce IDTs, which are reflected in Fig. 11(a) and 13.

The sensitivity analysis results for $C_3H_6$ are shown in Fig. 15. As observed previously, the reaction $N_2O(+M) = N_2 + O(+M)$ remains the most sensitive and strongly promoting reaction. In the updated model, the newly introduced reaction pathways, $C_3H_6 + N_2O = C_2H_5CHO + N_2$ and $C_3H_6 + N_2O = CH_3COCH_3 + N_2$, emerge as the second and third most sensitive and promoting reactions, respectively. Notably, the inclusion of these pathways suppresses the influence of all dominant inhibiting reactions, such as $C_3H_6 + H = C_2H_4 + CH_3$, $C_3H_6 + O = C_2H_4 + CH_2O$, etc. By introducing additional low-temperature oxidation channels, these new reactions substantially enhance the overall reactivity of the fuel system.

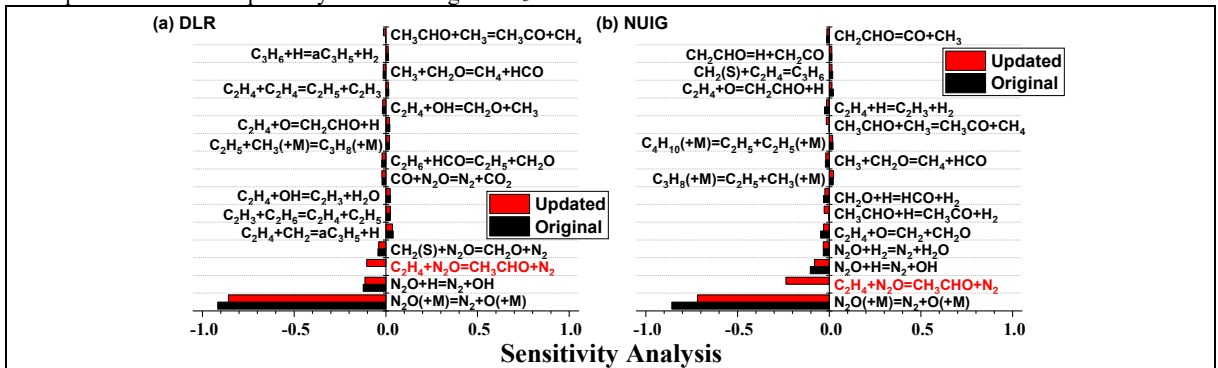

Figure 14. The sensitivity analysis for $C_2H_4$ at $P_C = 40$ bar, $T_C = 1000$K, equivalence ratio of 1 by (a) DLR and (b) NUIG model.

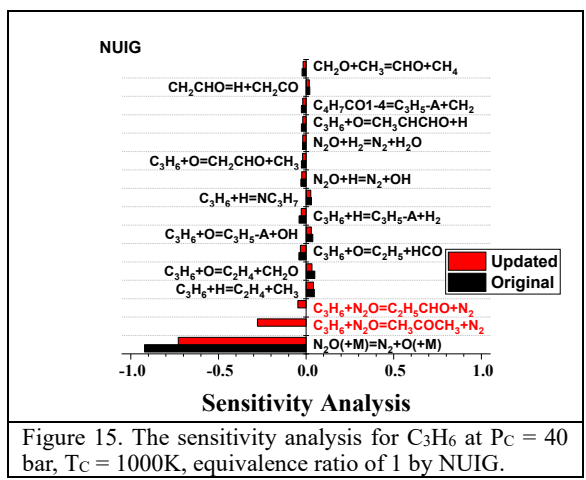

Figure 15. The sensitivity analysis for $C_3H_6$ at $P_C = 40$ bar, $T_C = 1000K$, equivalence ratio of 1 by NUIG.

### 3.4.4. Flux analysis

Flux analyses are performed using variable-volume simulations for $C_2H_4$ and $C_3H_6$ under the same conditions described in Section 3.4.3 (Figs. 14 and 15). The results, presented at 1% fuel consumption, indicate the distribution of each reactant's consumption among the different identified pathways. Upon incorporating the new reaction pathway into the model, substantial changes in flux distributions are observed.

Figure 16 shows the flux analysis results for $C_2H_4$ in the NUIG model at 1000K. With the addition of the O-atom transfer pathway from $N_2O$ to $C_2H_4$, the contributions from other $C_2H_4$ consumption pathways decrease, including certain reactions with reactivity-inhibiting effects, such as $C_2H_4 \rightarrow CH_2CHO$ and $C_2H_4 \rightarrow C_3H_6$, as shown in Fig. 14(b). In contrast, approximately 17.7% of $C_2H_4$ is consumed via the newly added pathway $C_2H_4 + N_2O = CH_3CHO + N_2$. This pathway is identified as the most sensitive and strongly promoting fuel-specific reaction in the system. Its major subsequent reaction, $CH_3CHO \rightarrow CH_3CO$, also ranks among the most promoting reactions in the sensitivity analysis presented in Section 3.4.3, consuming approximately 82% of $CH_3CHO$, the highest share among all $CH_3CHO$ consumption routes. Consequently, the O-atom addition from $N_2O$ to $C_2H_4$ and its subsequent oxidation sequence significantly shorten the ignition delay time. The results of $C_2H_4$ from DLR model (Fig. S3) exhibit a similar trend to the NUIG model, with quantitative differences observed: approximately 7.4% of $C_2H_4$ is consumed through the new pathway in DLR, leading to an increased $CH_3CHO$ production rate and accelerated $CH_3CO$ formation. These downstream processes directly contribute to the significant reduction in IDTs observed in Sections 3.4.1–3.4.2. In addition, the flux analysis results for $C_3H_6$ using the NUIG model show consistent results as $C_2H_4$, as shown in Fig. 17. The newly introduced reaction pathway, $C_3H_6 + N_2O = CH_3COCH_3 + N_2$ accounts for 20% of the total $C_3H_6$ consumption and represents the most sensitive, strongly promoting fuel-specific reaction. Another sensitive and promoting pathway, $C_3H_6 + N_2O = CH_3CH_2CHO + N_2$, contributes an additional 2.4% to the overall $C_3H_6$ consumption. Together, these reactions enhance the reactivity of the combustion system and lead to shorter IDTs.

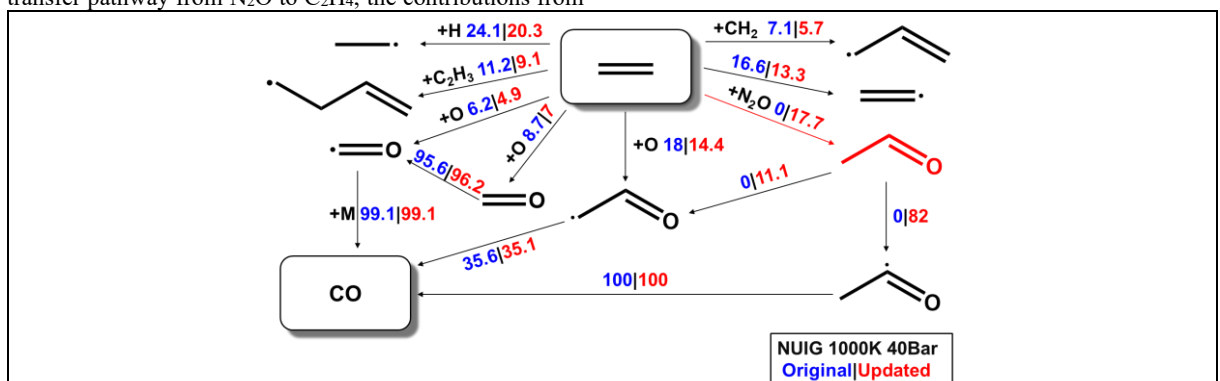

Figure 16. The flux analysis for $C_2H_4$ at $P_C = 40$ bar, $T_C = 1000K$, equivalence ratio of 1 by NUIG model. Fluxes are computed at 1% $C_2H_4$ consumption.

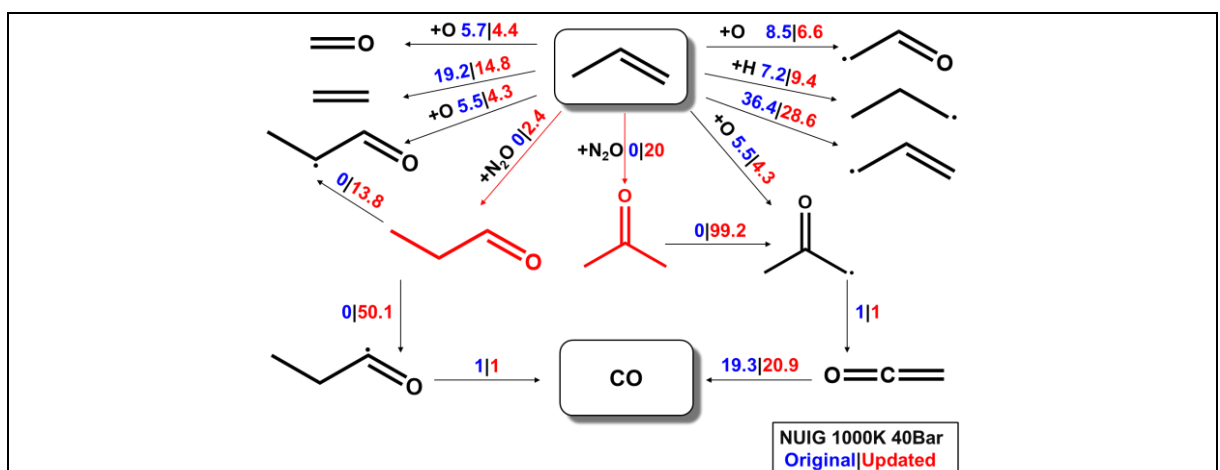

Figure 17. The flux analysis for $C_3H_6$ at $P_C = 40$ bar, $T_C = 1000K$, equivalence ratio of 1 by NUIG. Fluxes are computed at 1% $C_3H_6$ consumption.

## 4. Conclusions

This work characterizes new direct interaction pathways between $N_2O$ and unsaturated hydrocarbons, covering $C_2H_4$, $C_3H_6$, $C_2H_2$, and $C_3H_4$, through quantum chemistry calculations, comprehensive kinetic modeling, and experimental validations. These pathways have been missing from existing chemistry models and the corresponding rate coefficients have not been reported in the past. The key findings are as follows:

- The direct interaction pathways between $N_2O$ and unsaturated hydrocarbons proceed via O-atom addition reactions from $N_2O$ to unsaturated hydrocarbons proceed in multiple steps, which first form a five-member ring intermediate and eventually lead to the production of $N_2$ and different products depending on the hydrocarbon type and the reaction site for O-atom addition. The major difference between alkenes and alkynes lies in the decomposition pathway of the five-member ring intermediate, with the cleavage of N-C and N-O bonds and intramolecular H-atom transfer occurring in the same step for alkenes, while in different consecutive steps for alkynes.
- The distinctly different interaction pathways between alkenes and alkynes are further investigated via quantum mechanical analysis, which can be primarily attributed to the structural disparity of the five-membered ring intermediates with respect to the N-C bond length. Shorter and stronger N-C bonds are observed in alkyne intermediates than in alkene intermediates (i.e., ~1.480 Å in alkene intermediates vs. ~1.381 Å in alkyne intermediates), resulting in the higher energy required for N-C bond cleavage for alkyne intermediates.
- The rate coefficients of the direct interacting reactions are further determined and incorporated into four different kinetic models. Autoignition modeling results highlight the significant promoting impact of the new reactions on model reactivity for all models, with the O-atom addition reactions from $N_2O$ to unsaturated hydrocarbons emerging as the most sensitive fuel-specific promoting reaction. With the new reactions incorporated, the models' capability for autoignition simulation are extended to lower temperature ranges, with much better agreements obtained with experiments.
- Flux analysis reveals that the addition of the new pathways reduces fuel consumption via the conventional inhibiting pathways, while enables new dominant pathways for fuel consumption that lead to the formation of aldehydes and ketones, with both promoting reactivity.

**CRediT authorship contribution statement**

**Hongqing Wu**: Writing – review & editing, Writing – original draft, Validation, Software, Formal analysis, Conceptualization. **Guojie Liang**: Writing – review & editing, Writing –original draft, Validation, Software, Formal analysis, Conceptualization. **Tianzhou Jiang**: Writing – review & editing, Validation, Formal analysis, Data curation. **Fan Li**: Writing – review & editing, Validation, Formal analysis, Data curation. **Yang Li**: Writing – review & editing, Validation, Formal analysis, Data curation. **Rongpei Jiang**: Writing – review & editing, Validation, Formal analysis, Data curation. **Ruoyue Tang**: Writing – review & editing, Validation, Formal analysis, Data curation. **Song Cheng**: Writing – review & editing, Writing – original draft, Supervision, Resources, Project administration, Conceptualization.

**Acknowledgments**

The work described in this paper is supported by the Research Grants Council of the Hong Kong Special Administrative Region, China under 25210423 for ECS project funded in 2023/24 Exercise, the National Natural Science Foundation of China under 52406158, and the Natural Science Foundation of Guangdong Province under 2024A1515011486.

**Declaration of Competing Interests**

The authors declare that they have no known competing financial interests or personal relationships that could have appeared to influence the work reported in this paper.

**References**

[1] R. Tang, S. Cheng, Combustion Chemistry of Unsaturated Hydrocarbons Mixed with NOx: A Review with a Focus on Their Interactions, Energies 16 (2023) 4967.

[2] V. Zakirov, M. Sweeting, T. Lawrence, J. Sellers, Nitrous oxide as a rocket propellant, Acta Astronaut. 48 (2001) 353-362.

[3] D. Razus, Nitrous oxide: oxidizer and promoter of hydrogen and hydrocarbon combustion, Ind. Eng. Chem. Res. 61 (2022) 11329-11346.

[4] W. Pu, D. Sun, W. Fan, W. Pan, Q. Chai, X. Wang, Y. Lv, Cu-Catalyzed atom transfer radical addition reactions of alkenes with α-bromoacetonitrile, Chem. Commun. 55 (2019) 4821-4824.

[5] E. Zamir, Y. Haas, R. Levine, Laser enhanced addition reactions between hydrogen halides and unsaturated hydrocarbons: An information–theoretic approach, J. Chem. Phys. 73 (1980) 2680-2687.

[6] T.J. Preston, G.T. Dunning, A.J. Orr-Ewing, S.A. Vazquez, Direct and indirect hydrogen abstraction in Cl+ alkene reactions, J. Phys. Chem. A 118 (2014) 5595-5607.

[7] C. Zhou, A. Farooq, L. Yang, A.M. Mebel, Combustion chemistry of alkenes and alkadienes, Prog. Energ. Combust. 90 (2022) 100983.

[8] C. Naumann, T. Kick, T. Methling, M. Braun-Unkhoff, U. Riedel, Ethene/nitrous oxide mixtures as green propellant to substitute hydrazine: reaction mechanism validation, Int. J. Energetic Mater. Chem. Propul. 19 (2020).

[9] L. Werling, F. Lauck, D. Freudenmann, N. Röcke, H. Ciezki, S. Schlechtriem, Experimental investigation of the flame propagation and flashback behavior of a green propellant consisting of N2O and C2H4, J. Energy Power Eng. 11 (2017) 735-752.

[10] C. Janzer, S. Richter, C. Naumann, T. Methling, "Green propellants" as a hydrazine substitute: experimental investigations of ethane/ethene-nitrous oxide mixtures and validation of detailed reaction mechanism, CEAS Sp. J. 14 (2022) 151-159.

[11] F. Zhang, H. Chen, J. Feng, D. Zheng, Experimental investigation of auto-ignition of ethylene-nitrous oxide propellants in rapid compression machine, Fuel 288 (2021) 119688.

[12] R. Taylor, Safety and performance advantages of nitrous oxide fuel blends (NOFBX) propellants for manned and unmanned spaceflight applications, Safer Sp. Safer World 699 (2012) 67.

[13] X. Yang, X. Hong, W. Dong, Investigation on self-pressurization and ignition performance of nitrous oxide fuel blend ethylene thruster, Aerosp. Sci. Technol. 82 (2018) 161-171.

[14] S. Cheng, C. Saggese, S.S. Goldsborough, S.W. Wagnon, W.J. Pitz, Unraveling the role of EGR olefins at advanced combustion conditions in the presence of nitric oxide: Ethylene, propene and isobutene, Combust. Flame 245 (2022) 112344.

[15] Q. Peng, D. Huo, C.M. Hall. A comparison of neural network-based strategies for diesel engine air handling control. 2022 American Control Conf. (ACC); 2022: IEEE. p. 3031-3037.

[16] Q. Peng, D. Huo, C.M. Hall, Neural network-based air handling control for modern diesel engines, Proc. Inst. Mech. Eng., Part D 237 (2023) 1113-1130.

[17] S. Cheng, C. Saggese, S.S. Goldsborough, S.W. Wagnon, W.J. Pitz, Chemical kinetic interactions of NO with a multi-component gasoline surrogate: Experiments and modeling, Proc. Combust. Inst. 39 (2023) 531-540.

[18] S. Sawada, T. Totsuka, Natural and anthropogenic sources and fate of atmospheric ethylene, Atmos. Environ. 20 (1986) 821-832.

[19] J.C. Kramlich, W.P. Linak, Nitrous oxide behavior in the atmosphere, and in combustion and industrial systems, Prog. Energy Combust. Sci. 20 (1994) 149-202.

[20] R.L. Thompson, L. Lassaletta, P.K. Patra, C. Wilson, K.C. Wells, A. Gressent, E.N. Koffi, M.P. Chipperfield, W. Winiwarter, E.A. Davidson, Acceleration of global N2O emissions seen from two decades of atmospheric inversion, Nat. Clim. Change 9 (2019) 993-998.

[21] S. Johansson, G. Wetzel, F. Friedl-Vallon, N. Glatthor, M. Höpfner, A. Kleinert, T. Neubert, B.-M. Sinnhuber, J. Ungermann, Biomass burning pollution in the South Atlantic upper troposphere: GLORIA trace gas observations and evaluation of the CAMS model, Atmos. Chem. Phys. 2021 (2021) 1-23.

[22] R. Dodangodage, P. Bernath, C. Boone, M. Lecours, M. Schmidt, Atmospheric ethylene (C2H4) observations from the Atmospheric Chemistry Experiment Fourier Transform Spectrometer (ACE-FTS), J. Quant. Spectrosc. Radiat. Transfer, (2025) 109603.

[23] A. Trenwith, The kinetics of the oxidation of ethylene by nitrous oxide, J. Am. Chem. Soc., (1960) 3722-3726.

[24] F. Deng, Y. Pan, W. Sun, F. Yang, Y. Zhang, Z. Huang, Comparative study of the effects of nitrous oxide and oxygen on ethylene ignition, Energy Fuels 31 (2017) 14116-14128.

[25] T. Kick, J.H. Starcke, C. Naumann. Green propellant substituting hydrazine: investigation of ignition delay time and laminar flame speed of ethene/dinitrogen oxide mixtures. Proc. Eur. Combus. Meet. 2017; 2017.

[26] K.K. Yalamanchi, X. Bai, N.D. Fernando, A.S. Lua, S. Cheng, Y. Li, C.-W. Zhou, S.S. Goldsborough, S.M. Sarathy, From electronic structure to model application of key reactions for gasoline/alcohol combustion: Hydrogen-atom abstractions by CH3Ȯ radical, Combustion and Flame 252 (2023) 112742.

[27] C. Yang, J.-T. Chen, X. Zhu, X. Bai, Y. Li, K.K. Yalamanchi, S.M. Sarathy, S.S. Goldsborough, S. Cheng, H.J. Curran, From electronic structure to model application of key reactions for gasoline/alcohol combustion: Hydrogen-atom abstraction by CH3OȮ radicals, Proceedings of the Combustion Institute 39 (2023) 415-423.

[28] R. Tang, Y. Han, H. Chen, B. Qu, Y. Li, Z. Lu, Z. Xing, S. Cheng, Theoretical study of H-atom abstraction by CH3OȮ radicals from aldehydes and alcohols: Ab initio and comprehensive kinetic modeling, Combustion and Flame 259 (2024) 113175.

[29] V. Parmon, G. Panov, A. Uriarte, A. Noskov, Nitrous oxide in oxidation chemistry and catalysis: application and production, Catal. Today 100 (2005) 115-131.

[30] H. Wu, R. Tang, Y. Dong, X. Ren, M. Wang, T. Zhang, S. Cheng, H Atom Abstractions from C1–C4 Alcohols, Aldehydes, and Ethers by NO2: Ab Initio and Comprehensive Kinetic Modeling, J. Phys. Chem. A 129 (2025) 4724-4744.

[31] Z. Guo, H. Wu, R. Tang, X. Ren, T. Zhang, M. Wang, G. Liang, H. Guo, S. Cheng, Key Kinetic Interactions between NOX and Unsaturated Hydrocarbons: H Atom Abstraction from C3–C7 Alkynes, Dienes, and Trienes by NO2, J. Phys. Chem. A 129 (2025) 2584-2597.

[32] H. Wu, R. Tang, X. Ren, M. Wang, G. Liang, H. Li, S. Cheng, Understanding key interactions between NOX and C2-C5 alkanes and alkenes: the ab initio kinetics and influences of H-atom abstractions by NO2, Combust. Flame 272 (2025)

[33] Y. Zhao, D.G. Truhlar, The M06 suite of density functionals for main group thermochemistry, thermochemical kinetics, noncovalent interactions, excited states, and transition elements: two new functionals and systematic testing of four M06-class functionals and 12 other functionals, Theor. Chem. Acc. 120 (2008) 215-241.

[34] A.D. McLean, G.S. Chandler, Contracted Gaussian basis sets for molecular calculations. I. Second row atoms, Z= 11–18, J. Chem. Phys. 72 (1980) 5639-5648.

[35] W.J. Hehre, R. Ditchfield, J.A. Pople, Self—consistent molecular orbital methods. XII. Further extensions of Gaussian—type basis sets for use in molecular orbital studies of organic molecules, J. Chem. Phys. 56 (1972) 2257-2261.

[36] M.P. Andersson, P. Uvdal, New scale factors for harmonic vibrational frequencies using the B3LYP density functional method with the triple-ζ basis set 6-311+ G (d, p), J. Phys. Chem. A 109 (2005) 2937-2941.

[37] R.A. Kendall, T.H. Dunning Jr, R.J. Harrison, Electron affinities of the first-row atoms revisited. Systematic basis sets and wave functions, J. Chem. Phys. 96 (1992) 6796-6806.

[38] J.M. Martin, Ab initio total atomization energies of small molecules—towards the basis set limit, Chem. Phys. Lett. 259 (1996) 669-678.

[39] T.H. Dunning Jr, Gaussian basis sets for use in correlated molecular calculations. I. The atoms boron through neon and hydrogen, J. Chem. Phys. 90 (1989) 1007-1023.

[40] P. Zhang, S.J. Klippenstein, C.K. Law, Ab initio kinetics for the decomposition of hydroxybutyl and butoxy radicals of n-butanol, J. Phys. Chem. A 117 (2013) 1890-1906.

[41] F. Neese, ORCA 5.0.4, a quantum chemical program package; Max Planck Institute for Chemical Energy Conversion, Mülheim an der Ruhr, Germany, 2023. Available at: https://orcaforum.kofo.mpg.de.

[42] M.J. Frisch, G. Trucks, H.B. Schlegel, G. Scuseria, M. Robb, J. Cheeseman, G. Scalmani, V. Barone, G. Petersson, H. Nakatsuji, Gaussian 16, Gaussian, Inc. Wallingford, CT, 2016. Available at: https://gaussian.com.

[43] Y. Georgievskii, J.A. Miller, M.P. Burke, S.J. Klippenstein, Reformulation and solution of the master equation for multiple-well chemical reactions, J. Phys. Chem. A 117 (2013) 12146-12154.

[44] J.A. Miller, S.J. Klippenstein, Master equation methods in gas phase chemical kinetics, J. Phys. Chem. A 110 (2006) 10528-10544.

[45] Y. Georgievskii, S.J. Klippenstein, Strange kinetics of the C2H6+ CN reaction explained, J. Phys. Chem. A 111 (2007) 3802-3811.

[46] C. Eckart, The penetration of a potential barrier by electrons, Phys. Rev. 35 (1930) 1303.

[47] C. Zhou, Y. Li, U. Burke, C. Banyon, K.P. Somers, S. Ding, S. Khan, J.W. Hargis, T. Sikes, O. Mathieu, An experimental and chemical kinetic modeling study of 1, 3-butadiene combustion: Ignition delay time and laminar flame speed measurements, Combust. Flame 197 (2018) 423-438.

[48] W. Collins, S. Hochgreb, N. Swaminathan, J. Chen. Simulation of NOx formation in dilute H2/CO/N2–air diffusion flames using full and reduced kinetics. Proceedings of the European Combustion Meeting; 2007. Available at: http://www.me.berkeley.edu/gri_ mech/.

[49] Y. Wu, S. Panigrahy, A.B. Sahu, C. Bariki, J. Beeckmann, J. Liang, A.A. Mohamed, S. Dong, C. Tang, H. Pitsch, Understanding the antagonistic effect of methanol as a component in surrogate fuel models: A case study of methanol/n-heptane mixtures, Combust. Flame 226 (2021) 229-242.

[50] M.J. McNenly, R.A. Whitesides, D.L. Flowers, Faster solvers for large kinetic mechanisms using adaptive preconditioners, Proc. Combust. Inst. 35 (2015) 581-587.

[51] F. Karami, M. Vahedpour, Theoretical study on the gas phase reaction mechanism of acetylene with nitrous oxide, Struct. Chem. 24 (2013) 1513-1526.

[52] C. Naumann, T. Kick, T. Methling, M. Braun-Unkhoff, U. Riedel. Ethene/dinitrogen oxide-a green propellant to substitute hydrazine: Investigation on its ignition delay time and laminar flame speed, 26th ICDERS (2017).